# A single-sided linear synchronous motor with a high temperature superconducting coil as the excitation system

F. Yen[1, 2, 3], J. Li[1,2,3], S. J. Zheng[1,2,3], L. Liu[1,2,3], G. T. Ma[1, 2, 3], J. S. Wang[1, 3], S. Y. Wang[1, 3] and W. Liu[1,2,3]

[1]Applied Superconductivity Laboratory, Southwest Jiaotong University, Chengdu, Sichuan 610031, China
[2]State Key Laboratory of Traction Power, Southwest Jiaotong University, Chengdu, Sichuan 610031, China
[3]National Laboratory for Rail Transit, Chengdu, Sichuan 610031, China

Email: fei.h.yen@gmail.com

**Abstract:** Thrust measurements were performed on a coil made of $YBa_2Cu_3O_{7-\delta}$ coated conductor acting as the excitation system of a single-sided linear synchronous motor. The superconducting coil was a single pancake in the shape of a racetrack with 100 turns, the width and effective lengths were 42 mm and 84 mm, respectively. The stator was made of conventional copper wire. At 77 K and a gap of 10 mm, with an operating direct current of $I_{DC}$=30 A for the superconducting coil and alternating current of $I_{AC}$=9 A for the stator coils, thrust of 24 N was achieved. With addition of an iron core, thrust was increased by 49%. With addition of an iron back plate, thrust was increased by 70%.

**Keywords**: High temperature superconductivity, linear synchronous motor, superconducting coil, force measurements

## 1. Introduction:

High temperature superconducting (HTS) rotary motors have already shown great results when compared to conventional copper wound motors given that power output can be more than doubled and dimensions cut by half [1-3]. HTS generators are also popular nowadays with many being employed as wind energy generators [4]. It is therefore logical to explore whether HTS linear motors are practical for applications. Indeed, recent work, though all with BSSCO wire, has been done on HTS linear motors. Pina et al. have calculated and are in the process of obtaining experimental data on a combination of BSSCO wire wound armature and bulk YBCO piece mover [5]. Kim et al. have measured the thrust forces of a permanent magnet mover on a BSSCO wound three phase armature [6]. Oswald et al., have reported designs and AC loss data on a much larger scaled linear synchronous motors whose armature is also made of BSSCO [7]. On all three cases, the superconducting coils are first generation BSSCO wire and all use alternating current which creates AC losses. Their performance is also limited when operating at 77 K when a magnetic field is present. In the low temperature superconducting (LTS) regime, Atherton et al., reported nice results about their large scale linear synchronous motor project employing $Nb_3Sn$ wire and operating with direct current [8-10]. Our design is quite similar to that of Atherton's since we also conduct direct current through the superconducting coil and our end motive is to propel maglev cars [11], though we used second generation Zr doped $YBa_2Cu_3O_{7-\delta}$ (YBCO) coated conductor (CC) wire to form the excitation system and operate at 77 K. YBCO CC was chosen over BSSCO wire since the critical current of the former decays less under external magnetic field. Also, the bending diameter of BSSCO wire is over 50 mm whereas for YBCO CC is 11 mm which was critical in our case since the diameter of the innermost layer of the coil investigated was 12 mm. This is the first case where a linear synchronous motor has been made of HTS CC wire. We start with a small scale experiment so our excitation system only consisted of one superconducting coil which will eventually help us build a system consisting of four coils connected in series for the completion of a prototype. At 77 K, this coil is not able to generate high magnetic fields since the critical current density of the superconducting wire is limited at this temperature. Thus, an iron core and an iron back plate were employed to further improve magnetic interactions between the stator and the excitation coil. This report studies the effects of the coil when coreless and when having an iron core and an iron back plate on the thrust performance of the superconducting excitation system. The design in this report yields minimal initial costs and operational costs since the armature is non-superconducting and the superconducting excitation coil temperature needs only to be lowered to 77 K. Only 20 m of coated conductor wire was used to make one coil and if the maglev tracks already has an existing armature, all that is needed to be replaced is the old excitation system with a superconducting one with its coil width and coil pitch matched to that of the armature's.

## 2. Experimental Setup:

The excitation system consisted of one superconducting coil made of $YBa_2Cu_3O_{7-\delta}$ with Zr doping coated conductor wire material from SuperPower, Inc. The wire's more important parameters to our experiment were its self field critical current at 77 K of $I_c>85$ A, thickness of t=0.09 mm width of w=4.1 mm and rated bending radius of 5.5 mm. Other important parameters as well as its synthesis can be found elsewhere [12]. The superconducting coil was a single pancake coil in the form of a racetrack consisting of N=100 turns with the following parameters, inner radii of $a_{1a}$=6 mm and $a_{1b}$=37.5 mm, outer radii of $a_{2a}$=21 mm and $a_{2b}$=52.5 mm and height of 2b=4.1 mm (Fig. 1a). Kapton insulation tape was used to separate each turn layer. The coil form also had a racetrack shape and was made of Teflon with a cavity in the middle so an iron core can be inserted (Fig. 1b). The CC wire was bent onto the coil form 100 times to form the coil. The iron core and iron back plates also had a racetrack form with no laminations. The thickness and radii of the core were 10 mm and 4 mm and 35.5 mm, respectively. Similarly, the thickness and radii of the back-plate were 5 mm and 17 mm and 48.5 mm, respectively. Direct current was used to power the superconducting coil with an Agilent 6680A power supply.

The armature consisted of a conventional copper wound three phase single sided linear motor with a coil pitch of 42 mm. The number of turns was not made available by the manufacturer, but the magnetic field along the z direction at 6 A near 10 mm above its surface was measured to be 300-350 Gauss with a Hall Probe. Alternating current was used to power the armature from a variable frequency controller set at 31 Hz.

The superconducting coil was placed inside a home made liquid nitrogen cryostat and secured 10 mm above the armature comprising a single sided linear synchronous motor (SLSM) (Fig. 2). This 10 mm gap is the distance between the bottom of the coil and the top of the armature's case. The coil was allowed to only move along the x-axis, the direction of the traveling magnetic field generated by the armature. The thrust was measured by mechanically connecting the superconducting coil to a 100 Newton Interface force sensor mounted on the SCML-02 measurement setup described elsewhere [13]. The voltage output of the force sensor during calibrations and experiments were measured by an Agilent 34401A multimeter.

## 3. Results and Discussion:

The critical current of the racetrack coil was $I_C$=48 A at 77 K, down from its nominal value of $I_C$=85 A due to the coil's self field not due to bending since its bending diameter is of 11 mm. For the purposes of avoiding quenching of the coil magnet, direct current ranging only from $I_{DC}$=1 to 40 A was applied to the superconducting excitation system when subject to an AC magnetic field. The applied alternating current range to the armature varied from $I_{AC}$=1 to 9 A. Thrust measurements on four different cases were conducted: without a core and without a back-plate; with a back-plate only; with a core only; and with both a core and a back-plate.

Fig. 3 shows data obtained of the thrust with respect to the electrical phase of the armature at $I_{DC}$=30 A and $I_{AC}$=2 A, 3 A, 4.5 A, 6 A, 7.5 A and 9 A with no iron components. At $I_{AC}$=2 A, the peak thrust was 4.3 N. With an increase of either $I_{DC}$ or $I_{AC}$, peak thrust increased linearly (Fig. 3) to the highest measured value of 23.9 N when $I_{AC}$=9 A. A simplified analytical calculation can be used to verify the obtained thrust value from the following equation:

$$F_x = 2 L_y N I_{DCy} \times B_z(I_{AC}) \sin \theta \qquad (1)$$

$F_x$ is the thrust along the x-axis, $L_y$ the effective length of the coil along the y-axis, N the number of turns, $B_z(I_{AC})$ the average z component of the magnetic field generated by the armature at the height of the coil, θ the current angle and $I_{DCy}$ the y-direction component of the direct current through the superconducting coil. In essence, equation 1 without the factor of two on the right hand side is just the formula that represents the force experienced by a conducting wire with finite length carrying a current in the presence of a magnetic field. Equation 1 also does not contain a reluctance force component due to eddy currents generated from a change in $B_z(I_{AC})$ that lies on the x-y plane and the way our pancake coil is wound, each subsequent coated conductor layer is along the x-direction so the insulation layers are along the y-z plane effectively creating laminations that keep eddy current losses to a minimum. This is why $F_x$ with respect to time behaves almost identical to one single sinusoidal wave instead of a superposition of two sine waves of two frequencies encountered in typical cases. The normal force, $F_z$, did exhibit a reluctance force component dependent of frequency [14]. Thrust was found to increase linearly with either $I_{DC}$ or $I_{AC}$ since $B_z(I_{AC})$ also depends linearly on $I_{AC}$ (Fig. 4 and Fig. 5). The experimentally obtained $F_x$ values at different

$I_{DC}$ come to within 5 % of the calculated $F_x$ values from Equation 1, where N=100, $L_y$=0.0795 m and $B_z(I_{AC})$=0.035 T at $I_{AC}$=6 A. The same consistency was found on $F_x$ when setting $I_{DC}$=30 and varying $I_{AC}$. $L_y$ was calculated by adding all individual y components of length of each turn. For instance, from Fig. 1, the upper half of the outermost layer of the coil has a middle section of 63 mm that is straight along the y direction and two sections that can be equated to a semicircle of diameter 42 mm. The y component contribution to the length of this 42 mm diameter semicircle in the x-y plane is 21 mm along the y direction. Hence, the outermost turn has an $L_{y100}$=84 mm. $L_y$ was calculated for each turn and averaged.

When a back iron plate is positioned directly on top of a superconducting coil, the magnetic field lines behave as if a ferromagnetic mirror was present, so the height of the coil is virtually almost doubled but never reaching a factor of two [15]. The area and thickness needed for the back plate to almost double the magnetic field depends on the magnetic saturation of the material. In our assembly, the "mirror" having adequate thickness and area, was placed 1 mm above the coil due to a 1 mm wall thickness of the coil form. Furthermore, an almost doubling of the coil's magnetic field does not equate to an almost doubling of $F_x$ since the center point of the coil is no longer at 12 mm but rather at 14 mm away from the armature and $B_z(I_{AC})$ decreases with increasing distance. The end result was an increase of $F_x$ by 70% to 40.4 N when $I_{DC}$=30 A and $I_{AC}$=9 A (Fig. 4). Data curves collected for the cases of the coil having a back iron plate, iron core and both a core and back plate all look similar to data displayed in Fig. 3, the case when there is no core and no back plate, except with different amplitudes. For this reason, only their peak thrust values are reported (Fig. 4 and Fig. 5)

To operate an HTS SLSM at 77 K, it is not yet economically feasible to construct HTS coils that can generate 1 Tesla at 77 K. For this reason, we explored the option of employing iron cores. When an iron core was inserted into the racetrack superconducting coil's center, its critical current was increased slightly to $I_C$=49 A. Unlike LTS wire, YBCO coated conductor wire is highly anisotropic and in cases when the coils are too flat, such as in this case of a single pancake, its radial magnetic fields, $B_r$, become large. The pinning sites of the coated conductor are parallel to its surface so $I_C$ is more sensitive to $B_r$ than to $B_z$ [12]. With the addition of iron, the magnetic field lines were more concentrated to point along the $B_z$ direction and less along the $B_r$ direction, hence slightly increasing $I_C$. The single pancake coil without a core was too flat that its $I_C$ was limited by the coil's $B_r$. Thus, HTS superconducting coils for linear motors must have an optimized thickness, not too flat so that its $I_C$ is not being limited by $B_r$ and not too tall so that the top section of the coil has a weak magnetic force interaction with the armature. The optimization of superconducting coils comprising the excitation system for linear motors with respect to conventional armatures is discussed in another report [16].

With an iron core, at $I_{DC}$=30 A and $I_{AC}$=9 A, the maximum thrust was 35.7 N (Fig. 4), a 49% increase compared to the coreless and no back plate case. Analytically, the thrust can be estimated by the superposition of equation 1 and a second term that represents the iron that is proportional to the attractive force between two magnetic poles where its x-axis component

contribution is:

$$F_{x2} = B_{Fe} B_z (I_{AC}) A \sin \theta / 2 \mu_0 \quad (2)$$

$B_{Fe}$ is the saturation magnetic field inside the iron piece, $B_z(I_{AC})$ the magnetic field generated by the armature at the iron piece, A the area of the iron core in the x-y plane and $\mu_0$ the permeability of free space. The experimentally obtained $F_{x2}$ was lower than the result of the analytical expression since the areas of the CC wire coil pole and the armature's pole vary greatly creating large fringing fields. Even though $F_{x2}$ was lower compared to the $F_x$ generated by the superconducting parts of the coil, the presence of the iron core increased the overall thrust.

The highest maximum thrust value obtained was in the case of employing both an iron core and an iron back plate. At $I_{DC}$=30 A and $I_{AC}$= 9 A, a peak thrust of 46.1 N was measured (Fig. 4). The clearance gap was of 10 mm. Raising it to 15 mm will result in a lowering of $B_z(I_{AC})$ but $I_{DC}$ and $I_{AC}$ will both be able to be increased allowing for operation of the SLSM at 15 mm and produce nearly same thrust values. Also, in the dynamic setting, the SLSM will be locked in between magnetic fields so the AC fields will become very small, allowing for $I_{DC}$ to be increased closer to its $I_C$ value of 49 A.

## 4. Conclusions:

The thrust of a single pancake racetrack shaped superconducting coil acting as the excitation system of a single sided linear synchronous motor was measured and studied when not having an iron core, having an iron core and having an iron back plate. At $I_{DC}$=30 A and $I_{AC}$=6 A, the SLSM system generated a thrust of 24 N. Addition of an iron core and iron back plate increased the thrust by 49% and 70%, respectively. The effects of adding an iron core and iron back-plate on the performance of the SLSM were discussed. The current HTS SLSM design is the most practical one when applied to maglev systems in the sense that initial and operating costs are kept to a minimum; only the old conventional excitation system will need to be replaced by one made of CC wire and the costs for operation of the SLSM at 77 K will be low.

**Acknowledgements:**

This work was supported in part by the Traction Power State Key Laboratory of Southwest Jiaotong University grant number 2009TPL_Z02 and the Fundamental Research Funds for the Central Universities grant number SWJTU09CX056. We thank Professor David L. Atherton from Queens University for taking his time in answering many of our questions and providing us with invaluable suggestions.

**Figure Captions:**

Figure 1: (a) Single pancake racetrack shaped excitation coil with 100 turns and thickness of 4.1 mm made of Zr doped YBCO coated conductor (CC), all units in millimeters. (b) Picture of the CC coil impregnated in epoxy with an iron core in the center that is removable.

Figure 2: Single-sided linear synchronous motor comprised of a moving superconducting excitation coil (1) inside a liquid nitrogen Dewar (2) and a stationary conventional three phase armature (3). The coil was only able to move along the x direction and physically connected to a force sensor.

Figure 3: Thrust vs. current angle curves with $I_{DC}$=30 A through the HTS excitation coil and $I_{AC}$=2 A to 9 A through the armature of the HTS linear synchronous motor at 31 Hz.

Figure 4: Peak thrust experienced by the HTS coil when $I_{DC}$=30 A at different armature AC input values.

Figure 5: Peak thrust experience by the HTS coil when $I_{AC}$=6 A at different HTS coil DC input values.

Figure 1a:

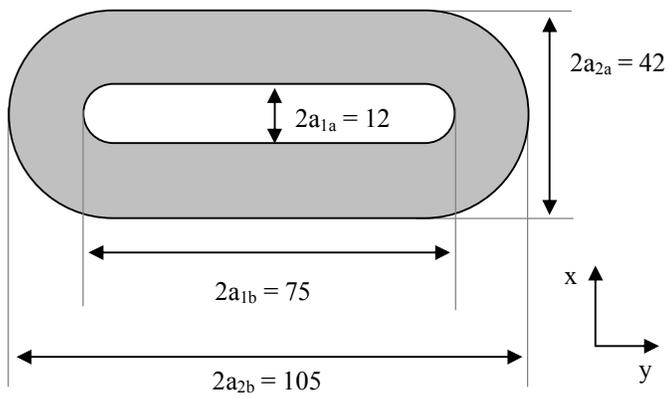

Figure 1b:

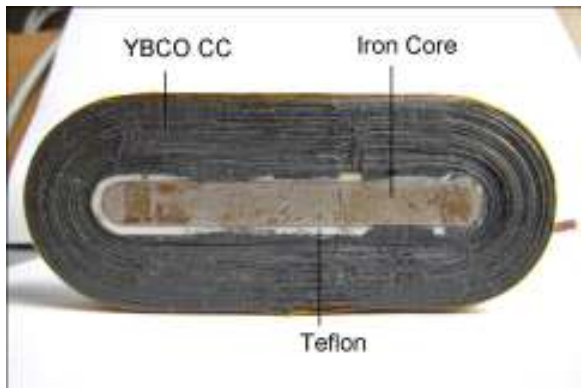

Figure 2:

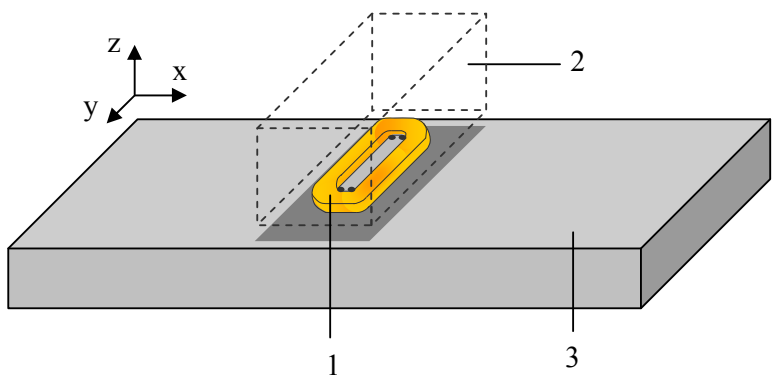

Figure 3:

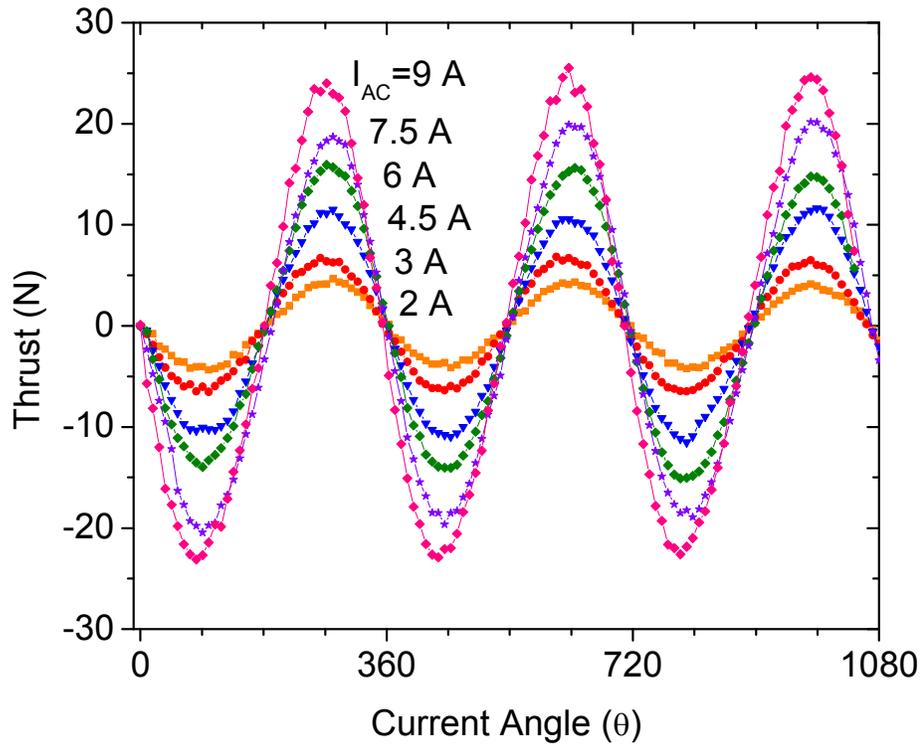

Figure 4:

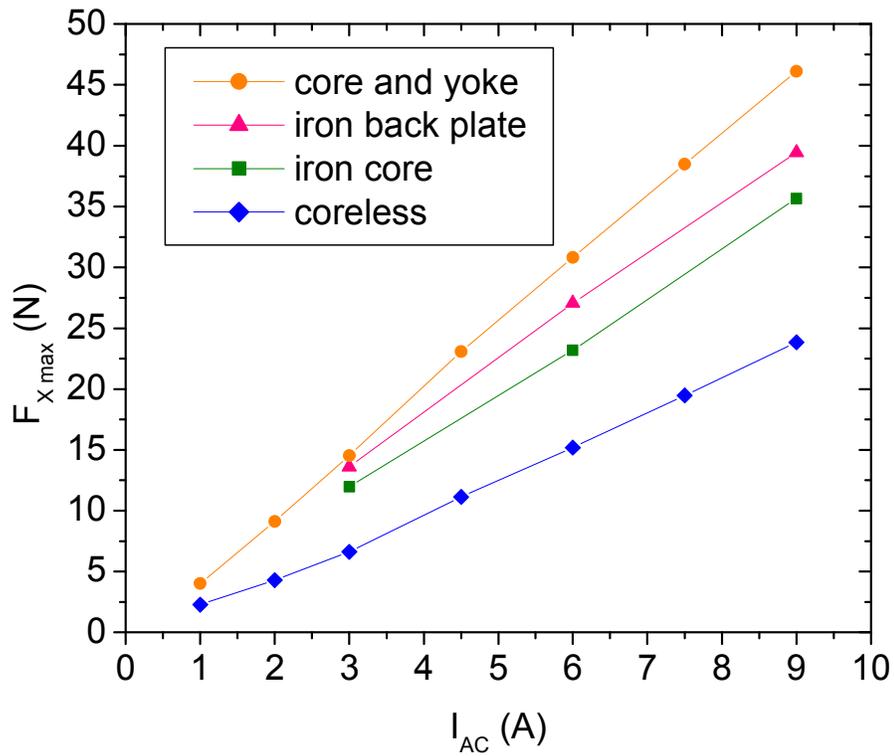

Fig. 5:

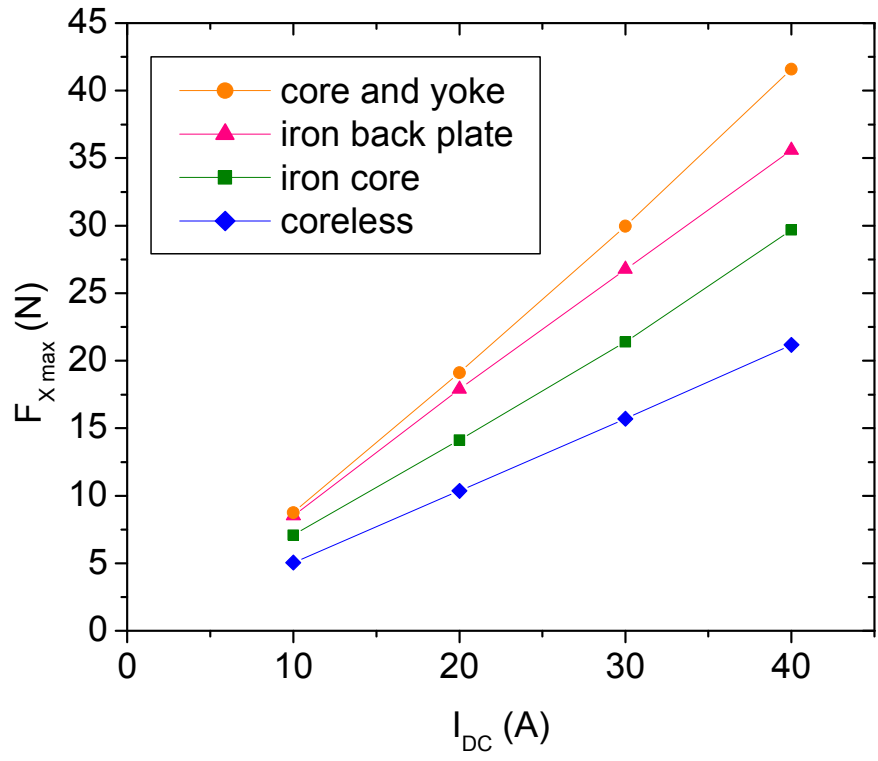